\documentclass{article}
\usepackage[preprint]{spconf}
\copyrightnotice{\copyright\ IEEE 2025}
\toappear{This work has been submitted to the IEEE for possible publication. Copyright may be transferred without notice, after which this version may no longer be accessible.}

\usepackage{amsmath,graphicx,hyperref}

\usepackage{cite}
\usepackage{enumitem}  
\usepackage{arydshln,booktabs,multirow,makecell}  
\usepackage{xcolor,colortbl}  
\usepackage{pifont,bbding}  
\usepackage{xspace}

\newcommand{\cmark}{\ding{51}\xspace}%
\newcommand{\xmark}{\ding{55}\xspace}%
\newcommand{\cellgray}{\cellcolor[HTML]{C0C0C0}}%

\title{AUV: Teaching Audio Universal Vector Quantization\\ with Single Nested Codebook}

\name{
Yushen Chen$^{1,2}$,
Kai Hu$^{3}$,
Long Zhou$^{3}$,
Shulin Feng$^{3}$,
Xusheng Yang$^{4}$,
Hangting Chen$^{3}$,
Xie Chen$^{1,2\dagger}$\thanks{$^\dagger$Corresponding Author.}}
\address{
    $^1$ X-LANCE Lab, Shanghai Jiao Tong University, China $^2$ Shanghai Innovation Institute, China\\
    $^3$ Tencent Hunyuan, China $^4$ Peking University, Shenzhen, China\\
}

\begin{document}
\ninept
\maketitle

\begin{abstract}
We propose AUV, a unified neural audio codec with a single codebook, which enables a favourable reconstruction of speech and further extends to general audio, including vocal, music, and sound. AUV is capable of tackling any 16 kHz mixed-domain audio segment at bit rates around 700 bps. To accomplish this, we guide the matryoshka codebook with nested domain-specific partitions, assigned with corresponding teacher models to perform distillation, all in a single-stage training. A conformer-style encoder-decoder architecture with STFT features as audio representation is employed, yielding better audio quality. Comprehensive evaluations demonstrate that AUV exhibits comparable audio reconstruction ability to state-of-the-art domain-specific single-layer quantizer codecs, showcasing the potential of audio universal vector quantization with a single codebook. The pre-trained model and demo samples are available at \url{https://swivid.github.io/AUV/}.
\end{abstract}

\section{Introduction}
\label{sec:intro}
Within the mainstream token-based language modeling paradigm for various speech and audio tasks \cite{audiolm,speartts,valle,uniaudio,moshi,qwen3omni}, a compact discrete representation that supports faithful reconstruction and distinguishable generation continues to grow in importance.
The key challenge of deriving neural audio codecs \cite{soundstream,encodec} capable of producing such discrete tokens lies in two aspects: reconstruction quality and generative capacity. 
The token rate and codebook size inherently constrain an audio codec's reconstruction quality. Relaxing the information bottleneck with a higher token rate or a larger codebook benefits reconstruction straightaway. However, larger token rates and codebook sizes add to the burden on downstream tasks, as larger models and training resources are required; otherwise, the generation performance would be sacrificed, since the audio sequence is longer and the modeling space becomes more complicated.

Neural audio codecs based on Residual Vector Quantization (RVQ) \cite{dac,flowdec,diffsoundstream} continue to push the limits of reconstruction fidelity and compactness. In contrast, recent works focused on developing single-layer quantizers \cite{singlecodec,wavtokenizer,bigcodec,lscodec,ts3codec,stablecodec,xcodec2,magicodec} are catching up, and further facilitating autoregressive modeling as the computation complexity and generation latency are lower.
On the other hand, researchers have carried out a considerable amount of exploration to encapsulate rich semantic information into discrete tokens produced by domain-specific codecs \cite{speechtokenizer,semanticodec,moshi,xcodec,mucodec,dualcodec,xytokenizer,yifan}. Still, relatively fewer investigations have been made from a unified audio view, including speech, music, and sound, especially with a desired single codebook paradigm \cite{unicodec}. 
Therefore, we frame our work here on the tokenization of general audio, providing practical and usable representations to boost unified generation tasks.

In this work, we mainly focus on the tokenization process itself, investigating and consolidating the single-layer quantization method to achieve faithful reconstruction of general audio with high perceptual quality:
\begin{itemize}[leftmargin=*]
    \item We introduce \textbf{AUV}, achieve \textbf{A}udio \textbf{U}niversal \textbf{V}ector quantization with a single nested codebook. AUV is with a domain-specific matryoshka codebook serving as a weak prior, and receives further semantic guidance through distillation from corresponding domain teacher models.
    \item We opt to use conformer as the neural audio codec encoder and decoder, with the time-frequency domain feature STFT as audio representation. We derive this effective model design through preliminary experiments, providing insights into the pros and cons of architecture backbone choices.
    \item Comprehensive evaluations show that AUV excels existing unified codecs with a single codebook in reconstruction. Results on the speech synthesis task also showcase the superiority of our method, initially revealing the potential in generation.
\end{itemize}

\section{Related Work}
\label{sec:prior}

\noindent
\textbf{Neural Audio Codecs}\quad A standard neural codec is an encoder-decoder model employing vector quantization \cite{vqvae} to the latent representation in the bottleneck. Discrete tokens from a finite codebook are typically extracted using the trained codec encoder and quantizer for downstream autoregressive generation tasks. Notably, most audio codecs are trained with the VQ-GAN framework \cite{vqgan}. From the research scope of unified single-codebook codecs, UniCodec \cite{unicodec} proposes a three-stage training with a rigid-split codebook and a domain-aware mixture-of-experts equipped encoder, surpassing its predecessor WavTokenizer \cite{wavtokenizer}, which directly uses a whole codebook for speech, music, and sound. However, UniCodec faces issues with complex multi-stage training, artifacts in reconstructed results, and a native deficiency in dealing with mixed-domain audio segments. Our work handles these challenges relatively better.
\\[-0.8em]

\noindent
\textbf{Semantic Audio Representation Learning}\quad Self-supervised learning (SSL) proficiency in speech-related tasks \cite{wav2vec2,hubert,wavlm} has helped a lot with incorporating semantic information into discrete tokens compressed by acoustic neural audio codecs, facilitating speech language modeling. SpeechTokenizer \cite{speechtokenizer} and Mimi \cite{moshi} perform distill-based semantic injection with SSL-based speech representation models. X-codec \cite{xcodec} adopts a dual-encoder structure, leveraging a pretrained speech SSL model as the semantic branch to work with another acoustic encoder. UniCodec \cite{unicodec} introduces a self-supervised, masked modeling approach in a second training stage to enhance semantic information. In this work, we further take advantage of music and general audio foundation models \cite{beats,muq} pre-trained with self-supervised learning to provide distillation signals with distinct domain characteristics.

\section{Method}
\label{sec:method}

\begin{figure}
\centering
\includegraphics[width=1\linewidth]{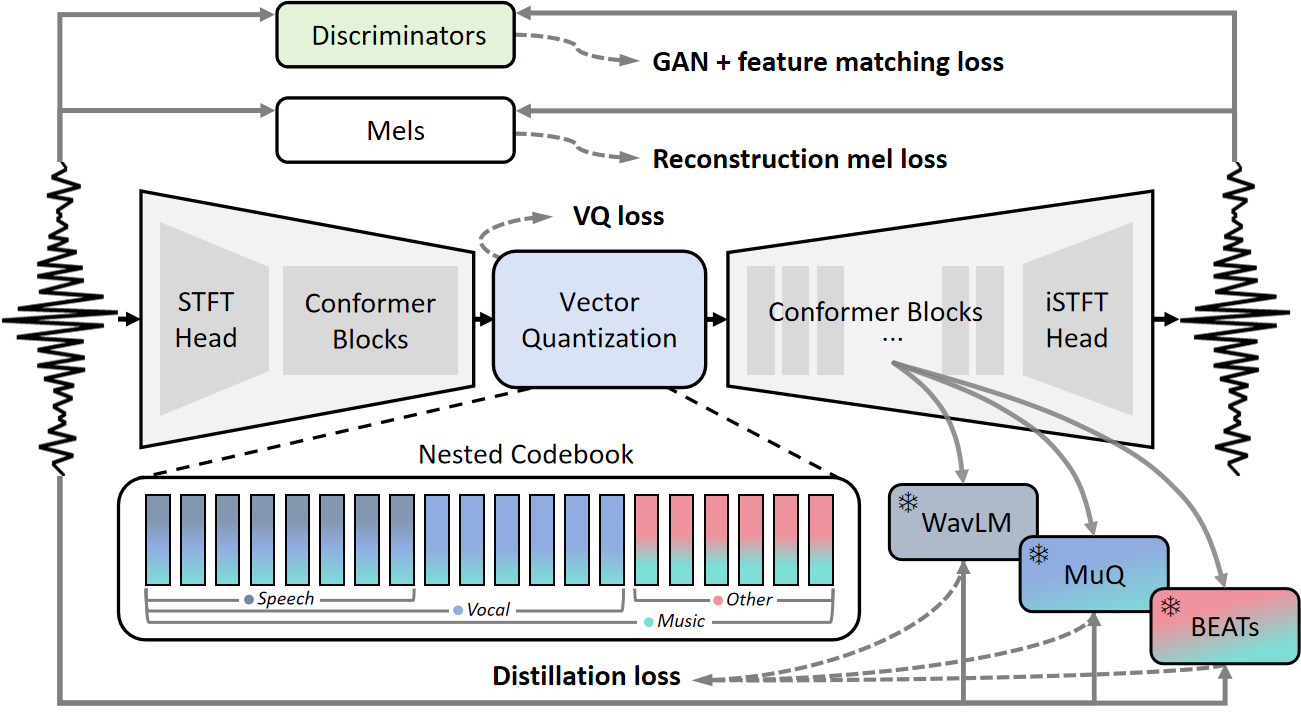}
\caption{Overview of AUV framework. During training, the input audio domain is made aware to the model, whereas it remains agnostic during inference, meeting the actual usage scenario.}
\label{fig:overview}
\end{figure}

In this section, we propose AUV, a single-layer acoustic audio codec incorporated with semantic distillation from a wide range of audio domains. We detail in the subsequent subsections.
\begin{itemize}[leftmargin=*]
    \item \textbf{Acoustic Neural Audio Codec}: The introduction of usual modules and the training procedure of an acoustic audio codec. Our model architecture and optimization strategies are also elaborated.
    \item \textbf{Nested Codebook for Universal Audio}: The design of a domain-specific matryoshka codebook for unified audio codecs.
    \item \textbf{Multi-domain Semantic Distillation}: The implementation of continuous representation distillation with teacher models to enrich the semantic information of our unified audio codec.
\end{itemize}

\subsection{Acoustic Neural Audio Codec}
\label{subsec:nac}
As illustrated in Fig. \ref{fig:overview}, AUV builds upon the framework established by existing acoustic codecs, namely EnCodec \cite{encodec}, DAC \cite{dac}, and so on. A single-codebook acoustic audio codec typically maintains an encoder-quantizer-decoder structure, the encoder transforms the input audio features into latent representations, the single-layer quantizer discretizes the latents into tokens, and the decoder finally reconstructs the waveform from the compressed discrete tokens.

Consistent with the findings of previous research \cite{audiodec}, our early experiments reveal that scaling the encoder yields minimal gains for reconstruction while scaling the decoder boosts reconstruction. Furthermore, we empirically found that the convolutional structure with local perceptual capability is essential to single-stage VQ-GAN training; and directly replacing all convolution layers with transformer blocks suffers severe performance loss or may require dedicated design, e.g., a multi-stage training pipeline in MagiCodec \cite{magicodec}.

Inspired by the success of Vocos \cite{vocos} and Conformer \cite{conformer} in various speech tasks, we propose to use conformer blocks as the encoder-decoder backbone and time-frequency domain feature STFT as the modeling target, with an STFT head to receive waveform input and an iSTFT head to recover the audio signal as output \cite{vocos}. We weight more parameters to the codec decoder for the reason mentioned above, in practice, using more layers for the decoder compared to the encoder. The Fourier transform is configured with a chosen hop length to obtain spectrum features, with the same token rate desired for discrete tokens after vector quantization. Thus, there are no learnable upsampling or downsampling modules in the structure. 
By comparing with the highly competitive baseline, which is VQ with Mimi \cite{moshi} encoder-decoder, we demonstrate the effectiveness of our model design. The details are provided in the experiments section.

The basic loss functions of AUV include quantizer loss, mel loss, adversarial loss, and feature matching loss, closely following BigCodec \cite{bigcodec}. The distillation-related loss will be introduced in Sec. \ref{subsec:distill}. A multi-period discriminator (MPD) from HiFi-GAN \cite{hifigan} and a multi-scale STFT (MS-STFT) discriminator are used for adversarial training, both in the way of BigCodec implementation. Notably, we incorporate an updated setting of FFT sizes \{206, 334, 542, 876, 1418, 2296\} from Stable-Codec \cite{stablecodec} used for the MS-STFT discriminator. This set of settings is originally demonstrated by Stable-Codec to mitigate periodic artifacts in the spectrum, where our own experiments also show that it significantly improves the perceptual quality of reconstruction, especially in speaker similarity.

\subsection{Nested Codebook for Universal Audio}
\label{subsec:codebook}
The design of a nested codebook for an all-in-one unified audio tokenizer comes from the intuition that the domains of speech, vocal, music, and sound are not strictly separated or totally distinct. In fact, we consider vocal data, which is singing without accompaniment, to be a superset of speech. Similarly, music samples can be regarded as encompassing vocals. 
Under this assumption, we believe that unblocking and sharing the codebook to a certain extent can enhance its utilization rate. Meanwhile, we expect the nested way as a weak prior for the model, after training on diverse data, to learn to spontaneously select and place more semantically inclined information into the codebook partition assigned to speech, and the more rhythmic information into the partition solely for the vocal domain. The same applies to music. In addition, sound segments without human voices can be regarded as existing relatively independently.

In terms of implementation, our nested codebook will be attributed to four mutually overlapping domains. With a total codebook size of 16384, indices from 0 to 4095 are allocated for the speech domain, 0 to 8191 for vocal, 0 to 16383 for music, and 8192 to 16383 for other non-human sound. When scaling data, we further extend 4096 entries to the nested codebook for speech-related domains (0-8191, 0-12287, 0-20479, 12288-20479, correspondingly for speech, vocal, music, and other). Noted that the allocation of regions is empirical and should be flexible to different encoder-decoder architectures. During training, the input audio domain is provided to the model, while in inference, the model is domain-agnostic. The audio codec model is expected to rely solely on the learned encoder and quantizer to select the appropriate token from the entire codebook.

Ablation experiments have been made between three different codebook design choices: the no-split codebook (a whole), the rigid-split codebook (UniCodec-style partitioning \cite{unicodec}), and the nested codebook (ours). The experimental results in Sec. \ref{subsec:ablation} show the superiority of our design, indicating a seamless integration of audio domains with AUV. And most importantly, index distribution probing statistics give a shred of strong evidence for our previous assumption that audio data from different domains are interrelated and mutually reinforcing.

\subsection{Multi-domain Semantic Distillation}
\label{subsec:distill}
The idea of performing semantic distillation is straightforward, and the methods are well-established. However, unlike previous works that only focus on speech, we propose to leverage all domain teacher models to distill a unified audio codec model.

For continuous representation distillation, the training objective is to simultaneously minimize the $L1$ distance and maximize the cosine similarity between the $D$-dimensional hidden representations of the teacher's $l$-th layer $\boldsymbol{h}_t^{(l)}$ and the student's distillation learner head's prediction $\boldsymbol{s}_t$. Formally, the distillation loss is defined as:
\begin{equation*}
    \begin{split}
        \mathcal{L}^{(l)}_{distill} = \sum_{t=1}^{T} \left[\frac{1}{D}\left\| \boldsymbol{s}_t - \boldsymbol{h}^{(l)}_t \right\|_1 - \log\sigma\left( \cos\left(\boldsymbol{s}_t, \boldsymbol{h}^{(l)}_t\right) \right) \right],
    \end{split}
    \label{eq:distill_loss}
\end{equation*}
where $T$ is the number of frames, $\sigma$ denotes sigmoid activation and $\cos (\cdot)$ indicates cosine similarity.

To be specific, we employ the average of the 13 to 24 layer representations of WavLM \cite{wavlm} as speech domain teacher; the average of the 5 to 10 layers of MuQ \cite{muq} as vocal and music domain teacher; and the average representation across all BEATs \cite{beats} layers as teacher for music and other (non-human sound) domains. The choice for MuQ is made based on a layer-wise investigation \cite{muqlayer}. The student's distillation learner head is on the 6th layer output hidden state of our AUV decoder, ablated and explained in Sec. \ref{subsec:ablation}.

\section{Experiments}
\label{sec:exprs}

\subsection{Experimental Setup}
\label{subsec:expr_setup}

\noindent
\textbf{Datasets}\quad We train AUV on approximately 120,000 hours of data. For speech, we use 95K-hour Emilia \cite{emilia} and LibriTTS \cite{libritts}. For vocal and music, the in-house dataset is around 20K hours. For audio, we filter Audio Set \cite{audioset} with labels to form a 4K-hour music set and an 800-hour non-human sound set. A 3K-hour mixed dataset is used for ablation, containing the 960h LibriSpeech \textit{train} set, 600h vocal/music each randomly selected from in-house data, and the 800h no-human subset from Audio Set. The speech, vocal, and audio reconstruction performances are evaluated on LibriSpeech \textit{test-clean} \cite{librispeech}, an in-house test set, and Audio Set \textit{eval}, respectively.
\\[-0.8em]

\noindent
\textbf{Pipeline}\quad All samples are resampled to 16 kHz with STFT hop length 320, resulting in a 50 Hz token rate. Segments longer than 12 seconds will be truncated during training. We use a global batch size of 128. The AdamW optimizer \cite{adamw} has a peak learning rate of 1e-4, where the scheduler linearly warmed up for 5K updates, cosine decayed for 500K updates, and remained constant over the rest of the training.
The hidden size of conformer blocks is 512, with an FFN multiply of 4. The codec encoder has 8 layers, and the decoder has 12 layers. We use 16384 and 20480 for codebook sizes (Sec. \ref{subsec:codebook}), and 8 for the factorized code dimension.
The Exponential Moving Average (EMA) \cite{ema} weights are used for inference.
\\[-0.8em]

\noindent
\textbf{Baselines}\quad In the scope of unified audio codecs with a single codebook, UniCodec \cite{unicodec} is the major baseline. Other leading single-layer quantizer models are also compared, including BigCodec \cite{bigcodec}, X-codec2 \cite{xcodec2}, and MagiCodec \cite{magicodec}. 
\\[-0.8em]

\noindent
\textbf{Metrics}\quad In reconstruction quality evaluation, for speech, we report Word Error Rate (WER) from a HuBERT-based \cite{hubert} ASR, Short Time Objective Intelligibility (STOI), Perceptual Evaluation of Speech Quality (PESQ), speaker similarity (SPK-SIM) from a WavLM-based \cite{wavlm} speaker verification model, and UTMOS \cite{utmos}; for vocal, music, and other sound, we report the unified automatic quality assessment scores from Audiobox Aesthetics \cite{audiobox_aesthetics}.

\begin{table}
\centering
\caption{Speech domain reconstruction evaluation results on LibriSpeech \textit{test-clean}. TPS stands for token per second.}
\label{tab:recons_librispeech}
\setlength\tabcolsep{3pt}
\resizebox{1\linewidth}{!}{
\begin{tabular}{lrcccccc}
\toprule
\textbf{Model} & \makecell{Codebook\\Size} & TPS & WER↓ & STOI↑ & \makecell{PESQ-\\WB↑} & \makecell{SPK-\\SIM↑} & \makecell{UT\\MOS↑} \\ 
\midrule
Ground Truth & -{ } & -            & 2.50 & 1.00 & 4.64 & 1.00 & 4.09 \\
\hline
DAC & 1024{ } & 50$\times$12       & 2.61 & 0.97 & 4.01 & 0.95 & 4.00 \\
BigCodec & 8192{ } & 80            & 3.63 & 0.94 & 2.68 & 0.84 & 4.11 \\
X-codec2 & 65536{ } & 50           & 3.20 & 0.92 & 2.43 & 0.82 & 4.12 \\
MagiCodec & 131072{ } & 50         & 4.25 & 0.92 & 2.54 & 0.77 & 4.17 \\
UniCodec & 16384{ } & 75           & 3.78 & 0.93 & 2.65 & 0.81 & 4.05 \\
UniCodec$_{w/\ id}$ & 4096{ } & 75 & 4.14 & 0.92 & 2.55 & 0.81 & 4.03 \\
\hline
\cellgray AUV & \cellgray 20480{ } & \cellgray 50 & \cellgray 3.64 & \cellgray 0.91 & \cellgray 2.40 & \cellgray 0.81 & \cellgray 4.09 \\
\bottomrule
\end{tabular}
}
\end{table}

\begin{table}
\centering
\caption{Other domain reconstruction results. Audiobox Aesthetics \cite{audiobox_aesthetics} scores include: Content Enjoyment (CE), Content Usefulness (CU), Production Complexity (PC), Production Quality (PQ).}
\label{tab:recons_other}
\setlength\tabcolsep{5pt}
\resizebox{1\linewidth}{!}{
\begin{tabular}{lcccccccc}
\toprule
\multirow{2.5}{*}{\textbf{Model}} & \multicolumn{4}{c}{\textbf{Vocal test set}} & \multicolumn{4}{c}{\textbf{Audio Set \textit{eval}}} \\ 
\cmidrule(lr){2-5} \cmidrule(lr){6-9}
& CE↑ & CU↑ & PC↑ & PQ↑ & CE↑ & CU↑ & PC↑ & PQ↑ \\ 
\midrule
Ground Truth & 5.69 & 6.04 & 3.44 & 6.81 & 4.52 & 5.73 & 4.10 & 6.33 \\
\hline
UniCodec     & 5.06 & 5.44 & 2.66 & 6.44 & 4.09 & 5.21 & 4.03 & 5.88 \\
AUV          & \textbf{5.90} & \textbf{6.16} & \textbf{3.33} & \textbf{6.85} & \textbf{4.27} & \textbf{5.40} & \textbf{4.08} & \textbf{6.02} \\
\bottomrule
\end{tabular}
}
\end{table}

\subsection{Reconstruction Evaluation}
\label{subsec:recons_eval}
Experimental results in Tab. \ref{tab:recons_librispeech} show that, as a unified acoustic audio codec, AUV is competitive with state-of-the-art domain-specific single-layer quantizer codecs in reconstruction quality. Our AUV has a lower token rate and modest codebook size, which has further advantages in usability. In Tab. \ref{tab:recons_other}, AUV comprehensively surpasses the previous leading model UniCodec, which further emphasizes that our design has great potential to serve as a general audio discrete representation to boost the development of downstream tasks.

\begin{figure}
\centering
\includegraphics[width=1\linewidth]{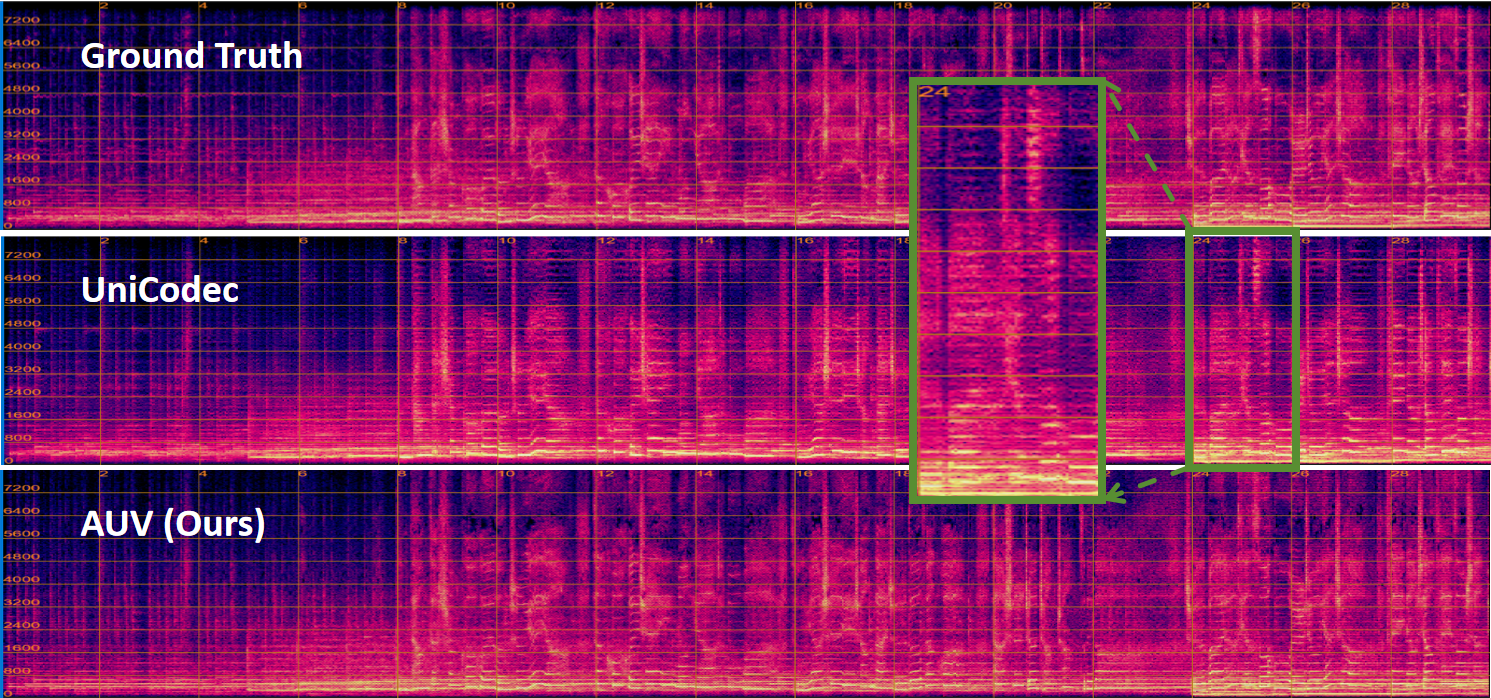}
\caption{Spectrograms of a snatch of music, including ground truth, and the reconstructed results by UniCodec and our AUV. UniCodec produces perceptible artifacts for audio samples other than speech.}
\label{fig:spectrogram}
\end{figure}

Compared to UniCodec, AUV has single-stage training, is able to tackle mixed-domain audio samples, and lacks artifacts. As UniCodec relies on a domain-aware mixture-of-experts equipped encoder with a rigid-split codebook, it cannot switch freely during inference, thus performs poorly when dealing with audio that contains information from different domains. Regarding artifacts, UniCodec has significant issues with over-smoothing and aliasing on waveforms, suffers great degradation in human auditory perception, and performs weakly in terms of audio sample' dynamics. Fig. \ref{fig:spectrogram} presents a set of spectrum comparisons as an example.

\begin{table*}[t]
\centering
\caption{Ablation of model architecture, disillation method, and multi-domain codebook design on LibriSpeech \textit{test-clean}.}
\label{tab:abl_modelarch}
\setlength\tabcolsep{4pt}
\resizebox{1\linewidth}{!}{
\begin{tabular}{@{}cccccccccccccccc@{}}


\toprule

\multirow{3.5}{*}{\textbf{ID}}
& \multirow{3.5}{*}{\makecell{Enc-Dec\\Arch.}}
& \multirow{3.5}{*}{\makecell{Trained\\Updates}}
& \multirow{3.5}{*}{\makecell{Data\\Domain}}
& \multirow{3.5}{*}{\makecell{Data\\Scale}}
& \multirow{3.5}{*}{\makecell{Distilled\\Dec Layer}}
& \multirow{3.5}{*}{\makecell{Codebook\\Type}}
& \multirow{3.5}{*}{\makecell{Codebook\\Size}}
& \multicolumn{5}{c}{\textbf{Reconstruction Quality}}
& \multicolumn{3}{c}{\textbf{Code Index Distribution Ratio}}
\\
\cmidrule(lr){9-13} \cmidrule(lr){14-16}
 & & & & & & & 
 & WER↓
 & STOI↑
 & \makecell{PESQ-\\WB↑}
 & \makecell{SPK-\\SIM↑}
 & \makecell{UT\\MOS↑}
 & Speech 
 & \makecell{Vocal-other-\\than-Speech}
 & Other
\\
\midrule

\textbf{(A0)} & Mimi & 700K & speech & 1K hrs & \xmark & - & 4096 & 5.23 & 0.91 & 2.26 & 0.73 & 4.14 & - & - & - \\
\textbf{(A1)} & AUV & 400K & speech & 1K hrs & \xmark & - & 4096 & 4.60 & 0.90 & 2.18 & 0.76 & 4.15 & - & - & - \\
\textbf{(A2)} & AUV & 300K & speech & 1K hrs & 4      & - & 4096 & 4.64 & 0.90 & 2.13 & 0.73 & 4.15 & - & - & - \\
\textbf{(A3)} & AUV & 200K & speech & 1K hrs & \cellgray 6      & - & 4096 & \cellgray 4.39 & \cellgray 0.90 & \cellgray 2.14 & \cellgray 0.76 & \cellgray 4.05 & - & - & - \\
\cdashline{1-16}\noalign{\vskip\belowrulesep}

\textbf{(B0)} & AUV & 1M & all & 3K hrs & \xmark & no split    & 16384 & 4.30 & 0.91 & 2.25 & 0.78 & 3.77 & 0.259 & 0.250 & 0.491 \\ 
\textbf{(B1)} & AUV & 1M & all & 3K hrs & \xmark & rigid split & 16384 & 4.21 & 0.91 & 2.31 & 0.79 & 3.86 & 0.322 & 0.214 & 0.464 \\
& UniCodec & - & all & 80K hrs          & -      & rigid split & 16384 & 3.78 & 0.93 & 2.65 & 0.81 & 4.05 & 0.321 & 0.109 & 0.571 \\
\textbf{(B2)} & AUV & 1M & all & 3K hrs & \xmark & \cellgray nested      & 16384 & \cellgray 3.99 & \cellgray 0.91 & \cellgray 2.35 & \cellgray 0.80 & \cellgray 3.89 & 0.371 & 0.189 & 0.441 \\
\cdashline{1-16}\noalign{\vskip\belowrulesep}

\textbf{(C0)} & AUV & 1M & all & 3K hrs   & \cellgray 6 & \cellgray nested & 16384 & 4.05 & 0.91 & 2.28 & 0.80 & 3.96 & \cellgray 0.382 & \cellgray 0.193 & \cellgray 0.424 \\
\textbf{(C1)} & AUV & 1M & all & 3K hrs   & 6 & nested & 20480 & 4.00 & 0.91 & 2.32 & 0.80 & 3.95 & 0.515 & 0.142 & 0.343 \\
\textbf{(C2)} & AUV & 1M & all & \cellgray 120K hrs & \cellgray 6 & \cellgray nested & \cellgray 20480 & \cellgray 3.64 & \cellgray 0.91 & \cellgray 2.40 & \cellgray 0.81 & \cellgray 4.09 & \cellgray 0.591 & \cellgray 0.109 & \cellgray 0.300 \\

\bottomrule
\end{tabular}
}
\end{table*}

\begin{table}
\centering
\caption{Speech domain generative capacity evaluation results on LibriSpeech-PC \textit{test-clean}. TPS stands for token per second.}
\label{tab:gen_tts}
\setlength\tabcolsep{2pt}
\resizebox{1\linewidth}{!}{
\begin{tabular}{cccccccc}
\toprule
\textbf{Used Codec} & \makecell{Multi-domain\\Distilled} & \makecell{Codebook\\Type} & \makecell{Codebook\\Size} & TPS & WER↓ & \makecell{SPK-\\SIM↑} & \makecell{UT\\MOS↑} \\ 
\midrule
Ground Truth & - & - & - & - & { }{ }1.86 & 0.69 & 4.09 \\
\hline
BigCodec & - & - & 8192 & 80 & 17.52 & \textbf{0.47} & 4.10 \\
X-codec2 & - & - & 65536 & 50 & 12.87 & 0.43 & 4.19 \\
UniCodec & - & rigid split & 16384 & 75 & 12.79 & 0.38 & 4.17 \\
\hline
\textbf{(B0)} & \xmark & no split    & 16384 & 50 & { }{ }5.45 & 0.43 & 4.15 \\
\textbf{(B1)} & \xmark & rigid split & 16384 & 50 & { }{ }6.26 & 0.43 & 4.20 \\
\textbf{(B2)} & \xmark & nested      & 16384 & 50 & { }{ }4.99 & 0.44 & 4.27 \\
\cdashline{1-8}
\textbf{(C0)} & \cmark & nested      & 16384 & 50 & { }{ }\textbf{4.51} & 0.44 & 4.26 \\
\textbf{(C2)} & \cmark & nested      & 20480 & 50 & { }{ }4.89 & 0.43 & \textbf{4.29} \\
\bottomrule
\end{tabular}
}
\end{table}

\subsection{Zero-shot TTS Results}
\label{subsec:res_tts}
To evaluate the generative capacity of AUV, we employ EmoVoice \cite{emovoice} as the text-to-speech (TTS) model to perform pure autoregressive stage training on the 1K-hour train set of LibriSpeech \cite{librispeech}, and evaluate on a test subset of LibriSpeech derived from F5-TTS \cite{f5tts}.

Results in Tab. \ref{tab:gen_tts} indicate the effectiveness of our nested codebook design and multi-domain distillation method, as (C0) and (C2) achieve much lower WER and higher UTMOS than models trained using other codec codes. (C0) has slightly better performance than (C2), and the reason for this can be attributed to the fact that the latter's larger codebook places higher demands on the modeling capabilities of downstream models (mentioned in Sec. \ref{sec:intro}), which may also be the cause of our failed training with MagiCodec extracted codes (its codebook size is 131072).

\subsection{Ablation Study}
\label{subsec:ablation}

\noindent
\textbf{Model Architecture}\quad Converged much faster, (A1) at a training update of 400K yields better WER, SPK-SIM, and UTMOS than the 700K-updated (A0) model. Thus, we opt to use conformer blocks with STFT as the backbone and conduct subsequent experiments.
\\[-0.8em]

\noindent
\textbf{Distillation Approach}\quad With a speech-only training, (A3) with a distillation learner on the 6th layer of the codec decoder stands out in WER compared to non-distilled and 4th-layer-distilled counterparts, (A1) and (A2) respectively. This result indicates that distillation is indeed beneficial for integrating semantic information into acoustic codecs, and it also shows that the reasonable selection of the student learner layer is crucial to maintain acoustic performance. Empirically, we found that a trivial probing test of attention rather than costly codec training can serve as a basis for selection; i.e., by discarding attention layers (of (A1)'s decoder) one by one and listening to the reconstructed speech, the layer that loses the most auditory quality due to discarding is thus found to be most correlated.
\\[-0.8em]

\noindent
\textbf{Codebook Design}\quad The reconstruction (in Tab. \ref{tab:abl_modelarch}) and generation (in Tab. \ref{tab:gen_tts}) capabilities of our proposed nested codebook are both verified ((B2) compared to no-split (B0) and rigid-split (B1)). The statistics of code index distribution highlight the rationality of considering the matryoshka codebook as a semantic prior, where (B2) attributes 37.1\% of codes to the speech-only partition in the codebook when inferencing on speech-domain input samples, 1.5 times in terms of possibility than a random guess (25\%). (C2)'s 59.1\% is likewise 1.5 times that of random (40\%, 8192 out of 20480). Moreover, the increase in the ratio from (B2) to (C0) adds to the effectiveness of multi-domain semantic distillation.
\\[-0.8em]

\noindent
\textbf{Data Scale}\quad Scaling the data has brought about a comprehensive gain in metrics ((C1) to (C2)). We are optimistic about further performance improvement if the data is further amplified.

\section{Conclusion}
\label{sec:conclu}
In this work, we introduced AUV, an advanced all-in-one speech, vocal, music, and sound codec with a single-layer quantizer at 50 Hz. Rich semantic information is seamlessly integrated into our unified acoustic codec. This is achieved using conformers with STFT heads as backbone, a domain-adaptive nested codebook design as an effective semantic prior, and a multi-domain distillation leveraging self-supervised learned teacher models. Our evaluations demonstrate that AUV outperforms existing unified neural codecs with a single codebook in reconstruction and generation, and is competitive with state-of-the-art domain-specific codecs.


\bibliographystyle{IEEEbib}
\bibliography{refs}

\begin{thebibliography}{10}

\bibitem{audiolm}
Zal{\'a}n Borsos et~al.,
\newblock ``Audio{LM}: a language modeling approach to audio generation,''
\newblock {\em IEEE/ACM TASLP}, 2023.

\bibitem{speartts}
Eugene Kharitonov et~al.,
\newblock ``Speak, read and prompt: High-fidelity text-to-speech with minimal supervision,''
\newblock {\em TACL}, 2023.

\bibitem{valle}
Sanyuan Chen et~al.,
\newblock ``Neural codec language models are zero-shot text to speech synthesizers,''
\newblock {\em IEEE TASLP}, 2025.

\bibitem{uniaudio}
Dongchao Yang et~al.,
\newblock ``Uni{A}udio: An audio foundation model toward universal audio generation,''
\newblock in {\em Proc. ICML}, 2024.

\bibitem{moshi}
Alexandre D{\'e}fossez, Laurent Mazar{\'e}, Manu Orsini, et~al.,
\newblock ``Moshi: a speech-text foundation model for real-time dialogue,''
\newblock {\em arXiv preprint arXiv:2410.00037}, 2024.

\bibitem{qwen3omni}
Jin Xu, Zhifang Guo, Hangrui Hu, et~al.,
\newblock ``{Qwen3-Omni Technical Report},''
\newblock {\em arXiv preprint arXiv:2509.17765}, 2025.

\bibitem{soundstream}
Neil Zeghidour, Alejandro Luebs, et~al.,
\newblock ``Sound{S}tream: An end-to-end neural audio codec,''
\newblock {\em IEEE/ACM TASLP}, 2021.

\bibitem{encodec}
Alexandre D{\'e}fossez, Jade Copet, et~al.,
\newblock ``High fidelity neural audio compression,''
\newblock {\em arXiv preprint arXiv:2210.13438}, 2022.

\bibitem{dac}
Rithesh Kumar, Prem Seetharaman, et~al.,
\newblock ``High-fidelity audio compression with improved {RVQGAN},''
\newblock {\em in Proc. NIPS}, 2023.

\bibitem{flowdec}
Simon Welker, Matthew Le, Ricky~TQ Chen, et~al.,
\newblock ``{FlowDec}: A flow-based full-band general audio codec with high perceptual quality,''
\newblock in {\em Proc. ICLR}, 2025.

\bibitem{diffsoundstream}
Yang Yang, Yunpeng Li, George Sung, et~al.,
\newblock ``{DiffSoundStream}: Efficient speech tokenization via diffusion decoding,''
\newblock {\em arXiv preprint arXiv:2506.22362}, 2025.

\bibitem{singlecodec}
Hanzhao Li, Liumeng Xue, Haohan Guo, et~al.,
\newblock ``Single-codec: Single-codebook speech codec towards high-performance speech generation,''
\newblock in {\em Proc. Interspeech}, 2024.

\bibitem{wavtokenizer}
Shengpeng Ji, Ziyue Jiang, Wen Wang, et~al.,
\newblock ``{WavTokenizer}: an efficient acoustic discrete codec tokenizer for audio language modeling,''
\newblock in {\em Proc. ICLR}, 2025.

\bibitem{bigcodec}
Detai Xin et~al.,
\newblock ``{BigCodec}: Pushing the limits of low-bitrate neural speech codec,''
\newblock {\em arXiv preprint arXiv:2409.05377}, 2024.

\bibitem{lscodec}
Yiwei Guo et~al.,
\newblock ``{LSCodec}: Low-bitrate and speaker-decoupled discrete speech codec,''
\newblock in {\em Proc. Interspeech}, 2025.

\bibitem{ts3codec}
Haibin Wu et~al.,
\newblock ``{TS3-Codec}: Transformer-based simple streaming single codec,''
\newblock in {\em Proc. Interspeech}, 2025.

\bibitem{stablecodec}
Julian~D Parker et~al.,
\newblock ``Scaling transformers for low-bitrate high-quality speech coding,''
\newblock in {\em Proc. ICLR}, 2025.

\bibitem{xcodec2}
Zhen Ye, Xinfa Zhu, Chi-Min Chan, et~al.,
\newblock ``Llasa: Scaling train-time and inference-time compute for llama-based speech synthesis,''
\newblock {\em arXiv preprint arXiv:2502.04128}, 2025.

\bibitem{magicodec}
Yakun Song, Jiawei Chen, et~al.,
\newblock ``{MagiCodec}: Simple masked gaussian-injected codec for high-fidelity reconstruction and generation,''
\newblock {\em arXiv preprint arXiv:2506.00385}, 2025.

\bibitem{speechtokenizer}
Xin Zhang et~al.,
\newblock ``{SpeechTokenizer}: Unified speech tokenizer for speech large language models,''
\newblock in {\em Proc. ICLR}, 2024.

\bibitem{semanticodec}
Haohe Liu et~al.,
\newblock ``{SemantiCodec}: An ultra low bitrate semantic audio codec for general sound,''
\newblock {\em IEEE JSTSP}, 2024.

\bibitem{xcodec}
Zhen Ye, Peiwen Sun, Jiahe Lei, et~al.,
\newblock ``Codec does matter: Exploring the semantic shortcoming of codec for audio language model,''
\newblock in {\em Proc. AAAI}, 2025.

\bibitem{mucodec}
Yaoxun Xu, Hangting Chen, et~al.,
\newblock ``{MuCodec}: Ultra low-bitrate music codec,''
\newblock {\em arXiv preprint arXiv:2409.13216}, 2024.

\bibitem{dualcodec}
Jiaqi Li, Xiaolong Lin, Zhekai Li, et~al.,
\newblock ``{DualCodec}: A low-frame-rate, semantically-enhanced neural audio codec for speech generation,''
\newblock in {\em Proc. Interspeech}, 2025.

\bibitem{xytokenizer}
Yitian Gong, Luozhijie Jin, et~al.,
\newblock ``{XY-Tokenizer}: Mitigating the semantic-acoustic conflict in low-bitrate speech codecs,''
\newblock {\em arXiv preprint arXiv:2506.23325}, 2025.

\bibitem{yifan}
Yifan Yang et~al.,
\newblock ``Towards universal speech discrete tokens: A case study for {ASR} and {TTS},''
\newblock in {\em Proc. ICASSP}, 2024.

\bibitem{unicodec}
Yidi Jiang et~al.,
\newblock ``{UniCodec}: Unified audio codec with single domain-adaptive codebook,''
\newblock {\em in Proc. ACL}, 2025.

\bibitem{vqvae}
Aaron Van Den~Oord, Oriol Vinyals, et~al.,
\newblock ``Neural discrete representation learning,''
\newblock {\em in Proc. NIPS}, 2017.

\bibitem{vqgan}
Patrick Esser, Robin Rombach, et~al.,
\newblock ``Taming transformers for high-resolution image synthesis,''
\newblock in {\em Proc. CVPR}, 2021.

\bibitem{wav2vec2}
Alexei Baevski, Yuhao Zhou, Abdelrahman Mohamed, and Michael Auli,
\newblock ``wav2vec 2.0: A framework for self-supervised learning of speech representations,''
\newblock {\em in Proc. NIPS}, 2020.

\bibitem{hubert}
Wei-Ning Hsu, Benjamin Bolte, Yao-Hung~Hubert Tsai, et~al.,
\newblock ``{HuBERT}: Self-supervised speech representation learning by masked prediction of hidden units,''
\newblock {\em IEEE/ACM TASLP}, 2021.

\bibitem{wavlm}
Sanyuan Chen et~al.,
\newblock ``Wavlm: Large-scale self-supervised pre-training for full stack speech processing,''
\newblock {\em IEEE JSTSP}, 2022.

\bibitem{beats}
Sanyuan Chen, Yu~Wu, Chengyi Wang, et~al.,
\newblock ``{BEATs}: Audio pre-training with acoustic tokenizers,''
\newblock in {\em Proc. ICML}, 2023.

\bibitem{muq}
Haina Zhu, Yizhi Zhou, Hangting Chen, et~al.,
\newblock ``{MuQ}: Self-supervised music representation learning with mel residual vector quantization,''
\newblock {\em arXiv preprint arXiv:2501.01108}, 2025.

\bibitem{audiodec}
Yi-Chiao Wu et~al.,
\newblock ``{AudioDec}: An open-source streaming high-fidelity neural audio codec,''
\newblock in {\em Proc. ICASSP}, 2023.

\bibitem{vocos}
Hubert Siuzdak,
\newblock ``Vocos: Closing the gap between time-domain and fourier-based neural vocoders for high-quality audio synthesis,''
\newblock in {\em Proc. ICLR}, 2024.

\bibitem{conformer}
Anmol Gulati, James Qin, Chung-Cheng Chiu, et~al.,
\newblock ``Conformer: Convolution-augmented transformer for speech recognition,''
\newblock in {\em Proc. Interspeech}, 2020.

\bibitem{hifigan}
Jungil Kong, Jaehyeon Kim, and Jaekyoung Bae,
\newblock ``{HiFi-GAN}: Generative adversarial networks for efficient and high fidelity speech synthesis,''
\newblock in {\em Proc. NIPS}, 2020.

\bibitem{muqlayer}
Yizhi Zhou, Haina Zhu, and Hangting Chen,
\newblock ``Layer-wise investigation of large-scale self-supervised music representation models,''
\newblock {\em arXiv preprint arXiv:2505.16306}, 2025.

\bibitem{emilia}
Haorui He, Zengqiang Shang, Chaoren Wang, et~al.,
\newblock ``Emilia: An extensive, multilingual, and diverse speech dataset for large-scale speech generation,''
\newblock in {\em Proc. SLT}, 2024.

\bibitem{libritts}
Heiga Zen et~al.,
\newblock ``{LibriTTS}: A corpus derived from librispeech for text-to-speech,''
\newblock in {\em Proc. Interspeech}, 2019.

\bibitem{audioset}
Jort~F Gemmeke et~al.,
\newblock ``{Audio Set}: An ontology and human-labeled dataset for audio events,''
\newblock in {\em Proc. ICASSP}, 2017.

\bibitem{librispeech}
Vassil Panayotov et~al.,
\newblock ``{LibriSpeech}: an {ASR} corpus based on public domain audio books,''
\newblock in {\em Proc. ICASSP}, 2015.

\bibitem{adamw}
Ilya Loshchilov and Frank Hutter,
\newblock ``Decoupled weight decay regularization,''
\newblock {\em arXiv preprint arXiv:1711.05101}, 2017.

\bibitem{ema}
Tero Karras et~al.,
\newblock ``Analyzing and improving the training dynamics of diffusion models,''
\newblock in {\em Proc. CVPR}, 2024.

\bibitem{utmos}
Takaaki Saeki et~al.,
\newblock ``{UTMOS}: Utokyo-sarulab system for voicemos challenge 2022,''
\newblock in {\em Proc. Interspeech}, 2022.

\bibitem{audiobox_aesthetics}
Andros Tjandra, Yi-Chiao Wu, et~al.,
\newblock ``{Meta Audiobox Aesthetics}: Unified automatic quality assessment for speech, music, and sound,''
\newblock {\em arXiv preprint arXiv:2502.05139}, 2025.

\bibitem{emovoice}
Guanrou Yang, Chen Yang, Qian Chen, et~al.,
\newblock ``{EmoVoice}: {LLM}-based emotional text-to-speech model with freestyle text prompting,''
\newblock in {\em Proc. ACM MM}, 2025.

\bibitem{f5tts}
Yushen Chen et~al.,
\newblock ``{F5-TTS}: A fairytaler that fakes fluent and faithful speech with flow matching,''
\newblock in {\em Proc. ACL}, 2025.

\end{thebibliography}

\end{document}